\documentclass[twocolumn,amssymb,reprint,nobibnotes,showpacs,superscriptaddress,aps,pra]{revtex4-2}
\usepackage{graphicx}
\usepackage[tbtags]{amsmath}
\usepackage{epsfig}
\usepackage{natbib}
\usepackage{upgreek}
\usepackage{multirow}
\usepackage{color}
\usepackage{hyperref}

\begin{document}

\title{Influence of ground-Rydberg coherence in two-qubit gate based on Rydberg blockade}

\author{Yangyang Liu}
\affiliation{State Key Laboratory of Magnetic Resonance and Atomic and Molecular Physics, 
Wuhan Institute of Physics and Mathematics, APM, Chinese Academy of Sciences , Wuhan 430071, China}
\affiliation{University of Chinese Academy of Sciences, Beijing 100049, China}

\author{Yuan Sun}
\email{yuansun@siom.ac.cn}
\affiliation{Key Laboratory of Quantum Optics and Center of Cold Atom Physics, Shanghai Institute of Optics and Fine Mechanics, Chinese Academy of Sciences, Shanghai 201800, China}

\author{Zhuo Fu}
\affiliation{State Key Laboratory of Magnetic Resonance and Atomic and Molecular Physics, 
Wuhan Institute of Physics and Mathematics, APM, Chinese Academy of Sciences , Wuhan 430071, China}
\affiliation{University of Chinese Academy of Sciences, Beijing 100049, China}

\author{Peng Xu}
\email{etherxp@wipm.ac.cn}
\affiliation{State Key Laboratory of Magnetic Resonance and Atomic and Molecular Physics, 
Wuhan Institute of Physics and Mathematics, APM, Chinese Academy of Sciences , Wuhan 430071, China}

\author{Xin Wang}
\affiliation{University of Chinese Academy of Sciences, Beijing 100049, China}
\affiliation{Key Laboratory of Quantum Optics and Center of Cold Atom Physics, Shanghai Institute of Optics and Fine Mechanics, Chinese Academy of Sciences, Shanghai 201800, China}

\author{Xiaodong He}
\affiliation{State Key Laboratory of Magnetic Resonance and Atomic and Molecular Physics, 
Wuhan Institute of Physics and Mathematics, APM, Chinese Academy of Sciences , Wuhan 430071, China}

\author{Jin Wang}
\affiliation{State Key Laboratory of Magnetic Resonance and Atomic and Molecular Physics, 
Wuhan Institute of Physics and Mathematics, APM, Chinese Academy of Sciences , Wuhan 430071, China}

\author{Mingsheng Zhan}
\affiliation{State Key Laboratory of Magnetic Resonance and Atomic and Molecular Physics, 
Wuhan Institute of Physics and Mathematics, APM, Chinese Academy of Sciences , Wuhan 430071, China}

\begin{abstract}
For neutral atom qubits, the two-qubit gate is typically realized via the Rydberg blockade effect, which hints about the special status of the Rydberg level besides the regular qubit register states. Here, we carry out experimental and theoretical studies to reveal how the ground-Rydberg coherence of the control qubit atom affects the process of two-qubit Controlled-Z ($C_Z$) gate, such as the commonly used ground-Rydberg $\pi$-gap-$\pi$ pulse sequence originally proposed in Phys. Rev. Lett. \textbf{85}, 2208 (2000). We measure the decoherence of the control qubit atom after the $\pi$-gap-$\pi$ pulses and make a direct comparison with the typical decoherence time $\tau_{gr}$ extracted from Ramsey fringes of the ground-Rydberg transition. In particular, we observe that the control qubit atom subject to such pulse sequences experiences a process which is essentially similar to the ground-Rydberg Ramsey interference. Furthermore, we build a straightforward theoretical model to link the decoherence process of control qubit subject to $C_Z$ gate $\pi$-gap-$\pi$ pulse sequence and the $\tau_{gr}$, and also analyze the typical origins of decoherence effects. Finally, we discuss the $C_Z$ gate fidelity loss due to the limits imposed by the ground-Rydberg coherence properties and prospective for improving fidelity with new gate protocols.
\end{abstract}
\maketitle

\section{{Introduction}}
\label{Introduction}

For the neutral atom qubit platform, the characterization of the qubit quality is an essential topic in its applications in quantum computation and quantum simulation. Qubits serve as the building blocks of quantum computation, and the coherence of the qubit register states $|0\rangle, |1\rangle$ is one of the most essential characteristics to ensure the performances of qubits. The study of coherence has constantly attracted intense interests and efforts across various qubit platforms, such as neutral atoms, trapped ions \cite{T.Monz2011, C.J.Ballance2016}, superconductors \cite{R.Barends2014, C.Song2017}, quantum dots and so on. On the other hand, the two-qubit controlled-NOT (CNOT) gate, as the crucial ingredient for universal quantum computing, typically involves interaction with one or more extra auxiliary states. Despite the overwhelming attention to the subject of coherence in general, the subtle issue of coherence between qubit register states and the auxiliary states does not often receive the same level of scrutiny.

The neutral atom qubit platform has made consistent improvement in quantum information processing recently, such as high fidelity up to F=99.99\% of single qubit gates \cite{Xia_2015, K.Maller2015, Y.Wang2016, ChengSheng2018}, neutral atom arrays containing up to hundreds of atoms \cite{M.Endres2016, D.Barredo2016, Mello2019} and long coherence time of qubit register states as much as hundreds of millisecond of single atom \cite{J.Yang2016, Guo2020}. In particular, the fidelity of two-qubit gates based on Rydberg blockade have been consistently improved to more than 0.9 currently \cite{Isenhower2010, K.Maller2015, Y.Zeng2017, T.M.Graham2019, Levine2019, Madjarov2020}, by suppressing the phase noise on the Rydberg excitation lasers \cite{Sylvain2018, Levine2019} and by cooling atoms down to ground vibrational states in tweezers \cite{Madjarov2020}. However, the experimental fidelities of two-qubit CNOT gate and Bell state are still significantly lower than theoretical predictions. Some previous works have mentioned the relation between the single-atom ground-Rydberg coherence and the two-qubit gate process \cite{T.M.Graham2019}, but a convincing and complete experimental analysis is still missing. In other words, a thorough investigation will not only enhance the conceptual understanding of the neutral atom qubit platform, but will also be of help to the improvement of two-qubit gate fidelities into the next stage. Moreover, such a study will also provide insights to the problem how to practically improve the atom-photon gate fidelity based on the Rydberg blockade effect \cite{PhysRevA.93.040303, PhysRevA.94.053830, OPTICA.5.001492}.

In this article, based on the widely used CNOT gate via Rydberg blockade which was first proposed in Ref. \cite{D.Jaksch2000}, we carry out experimental and theoretical studies to reveal how the ground-Rydberg coherence of the control qubit atom affects the process of two-qubit $C_Z$ gate. The motivation is to demonstrate the influence of single qubit atom's ground-Rydberg coherence in two-qubit gate based on Rydberg blockade. In particular, we have designed and realized an experiment to verify the close relationship between the single-atom ground-Rdyberg decoherence time $\tau_{gr}$ and the fidelity of Bell state prepared by H-$C_Z$ type CNOT gate.

The rest of this article is organized as follows. In Sec. \ref{sec:Experiment setup} we present the basic descriptions about the experimental setup, together with the ground state operation of a single qubit. In Sec. \ref{sec:Coherent control and coherence of single qubit} we demonstrate coherent control of qubits and measure the decoherence time both of ground state and ground-Rydberg state. In Sec. \ref{sec:characterization the coherence of control qubit in a C-NOT gate}, we experimentally verify the close relationship between $\tau_{gr}$ and fidelity of Bell state prepared by H-$C_Z$ type CNOT gate in experiment. In Sec. \ref{sec:theory}, the influence of $\tau_{gr}$ on fidelity is explained theoretically. In Sec. \ref{sec:Discussion}, we summarize the major findings and discuss methods to improve fidelty of two-qubit gate, where the prospects for new gate protocols are also included.

\section{Experiment setup}
\label{sec:Experiment setup}

\begin{figure}[htbp]
\includegraphics [width=0.48\textwidth]{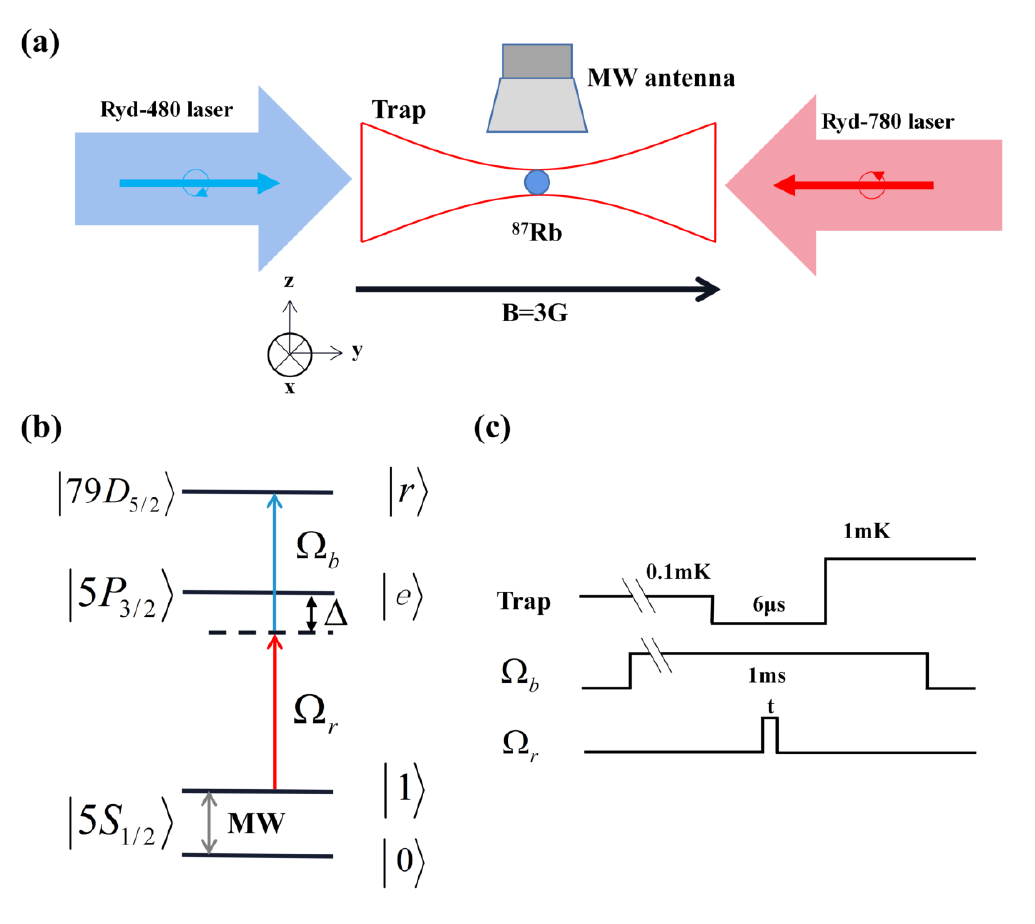}
\caption{(color online) (a) Simplified illustration of experiment setup. A microwave antenna is located 10 mm away from the dipole trap to coherently drive the transition between $|0\rangle$ and $|1\rangle$. (b) Qubit level scheme of $^{87}$Rb with $|0\rangle=5S_{1/2},F=1,m_F=0$, $|1\rangle=5S_{1/2},F=2,m_F=0$ and $|r\rangle=79D_{5/2}, m_j=5/2$. (c) Rydberg excitation pulse sequence. The 480 nm laser is applied during the entire process for about 1 ms. Note that in every experiment cycle reported in this article, the dipole trap are turned off during the time interval of 6--10 $\mu$s while applying the Rydberg excitation pulses. After excitation, we increase the trap depth to a large value ($\sim$ 1 mK) to detect Rydberg atoms. Only atoms in the ground state are recaptured by the trap, whereas those left in the Rydberg state are pushed away or ionized by the trap beam.}
\label{fig:fig1}
\end{figure}

The schematic experimental setup is shown in Fig. \ref{fig:fig1}(a). A linear polarized 830 nm laser is strongly focused into the vacuum chamber to form an optical dipole trap using a high numerical aperture (NA) lens (Qioptiq Optem 28-20-45-000).
The optical paths are specially designed and carefully adjusted to minimize the abbreviations caused by the 3 mm thick windows, and thus the beam waist of the dipole laser at the focal plane is optimized to 1.2 $\mu$m.
Individual $^{87}$Rb atoms are loaded from a magneto-optical trap (MOT) into this dipole trap of 0.6 mK trap depth.
Then, the polarization gradient cooling and adiabatic cooling by lowering the trap depth from 0.6 mK to 0.05 mK are applied to cool single-atoms down to 5.2 $\mu$K, as measured by release and capture method \cite{PhysRevA.78.033425}.
We use a bias magnetic field of 3 G along the Y axis to define the quantization axis and prepare atoms into $|1\rangle=|5S_{1/2},F=2,m_F=0\rangle$ by optical pumping with the efficiency $>99\%$.

The coherent transition between $|0\rangle= |5S_{1/2},F=1,m_F=0\rangle$ and $|1\rangle$ is realized by 6.8 GHz microwave, and the coherent excitation from $|1\rangle$ to the Rydberg states $|r\rangle = |79D_{5/2}, m_j = 5/2 \rangle$ is realized by a two-photon transition. We use 480 nm laser ($\Omega_b$) with $\sigma^+$ polarization and 780 nm laser ($\Omega_r$) with $\sigma^+$ polarization which is red-detuned by 5.7 GHz with respect to the transition from 5S$_{1/2}$ to 5P$_{3/2}$. To minimize the Doppler effect caused by atoms' motion, the two-photon excitation is settled in a counter-propagating geometry as shown in Fig. \ref{fig:fig1}(a). As mentioned in our recent work \cite{Zeng2018}, the Rydberg excitation lasers are frequency stabilized by the Pound-Drever Hall (PDH) method to a tunable reference cavity with finesse 58800 at 780nm and 91100 at 960nm. The laser linewidths are optimized to be less than 1 kHz and the long-term frequency drifts are limited to be less than 22 kHz for 10 hours. Moreover, to further reduce the laser phase noise, we follow the approaches of \cite{H.Levine2018, N.Akerman2015, L.Gerster2015} to use the transmitted lights from the high fineness cavity to install an extra injection lock to laser diodes, which inherits the same spectral properties. In this way, the excitation lasers' phase noise are suppressed by almost 3 orders below 1 MHz.

The Rydberg excitation process follows the time sequence illustrated in Fig. \ref{fig:fig1}(c). The dipole trap is turned off during the excitation and turned on with 1 mK trap depth to enable efficient Rydberg state detection.  Due to finite temperature,  atoms may move outside of the trapping volume, resulting in atom loss. The probability of atom loss in our setup is $1\%$ to $5\%$ depending on the trap-off time.
The $|0\rangle, |1\rangle$ states of qubit atoms are detected by applying a circular polarized laser resonant with $5S_{1/2}, F=2 \leftrightarrow 5P_{3/2}, F=3$ to push out atoms in $F=2$ and left atom in $F=1$ alone. We have measured the state selection efficiency to be larger than $99\%$ for atoms in either state. All data points here are extracted from 100 to 300 repeats of the experiment cycles.

\section{Coherent control and coherence of single qubit atom}
\label{sec:Coherent control and coherence of single qubit}

\begin{figure}[htbp]
\centering
\includegraphics [width=0.48\textwidth]{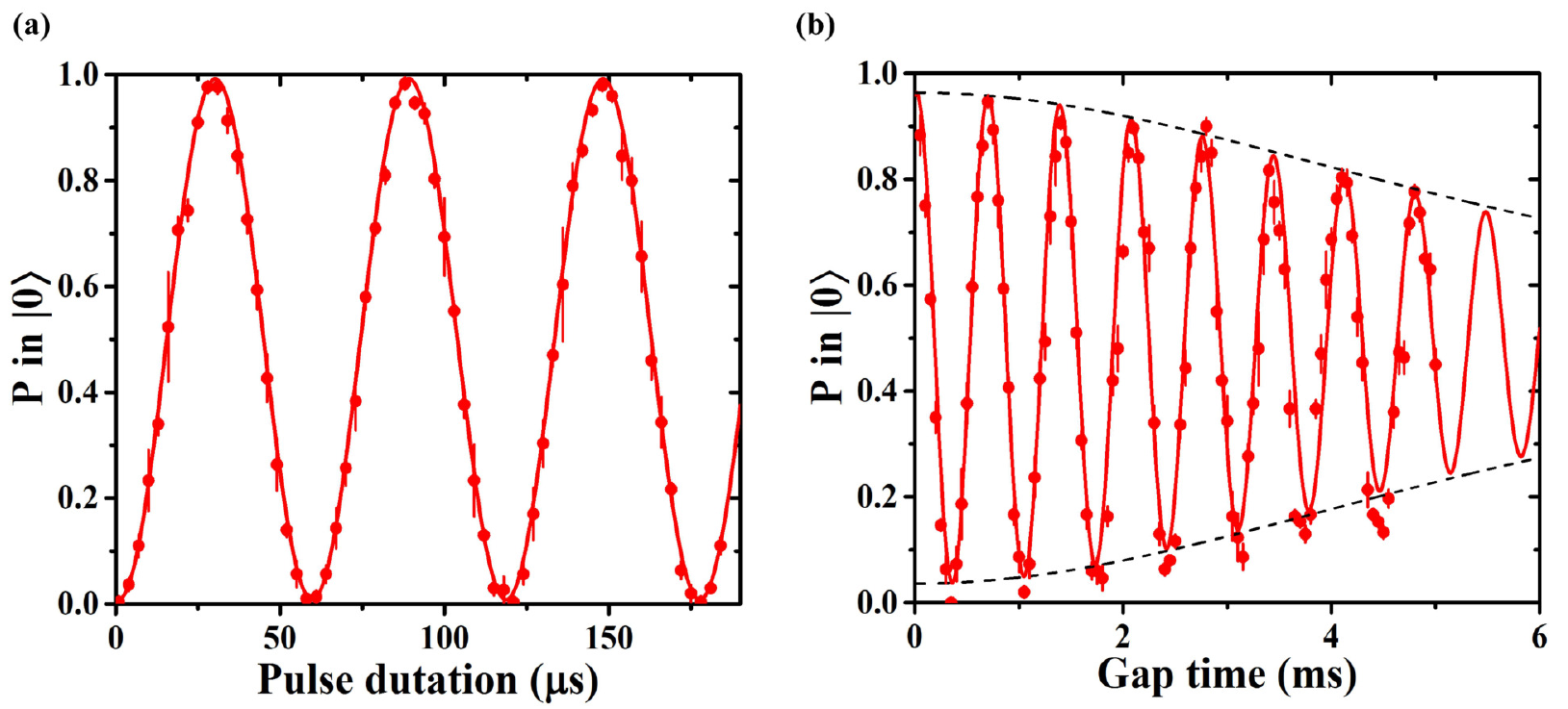}
\caption{(color online) (a) Rabi oscillation between $|0\rangle$ and $|1\rangle$ with an effetive Rabi frequency $\Omega = 2\pi \times 33.9$ kHz, corresponding to a ground $\pi$ pulse time $t_\pi = 29.5 \mu$s. (b) The Ramsey fringe between $|0\rangle$ and $|1\rangle$ with fitted $T_{2,gg}^*=7.2 \pm 0.6$ ms using $P_\text{Ramsey}(t)=B + \alpha(t,T_{2,gg}^*)A\cos[\delta' t + \kappa(t, T_{2,gg}^*) + \phi]$. The envelope is the function of $B \pm A\alpha(t,T_{2,gg}^*)$ \cite{S.Kuhr2005}.}
\label{fig:fig2}
\end{figure}

First, we study the coherence between $|0\rangle$ and $|1\rangle$ of the atomic qubit. The Rabi flopping is demonstrated by recording the population in $|0\rangle$ as a function of microwave pulse duration, and a typical trace of data is shown in Fig. \ref{fig:fig2}(a). The contrast of the Rabi oscillation is less than ideal due to state preparation and measurement (SPAM) errors ($<2\%$), and the transfer efficiency between $|0\rangle$ and $|1\rangle$ is above $99\%$, as explained in more details in \cite{ChengSheng2018}.
We apply a Ramsey sequence to measure the coherence time $T_{2,gg}^*$ between $|0\rangle$ and $|1\rangle$ , as presented in \ref{fig:fig2}(b). By fitting the data with $P_\text{Ramsey}(t)=B + \alpha(t, T_{2,gg}^*)A\cos[\delta' t + \kappa(t, T_{2,gg}^*) + \phi]$ where $\alpha(t,T_{2,gg}^*)=[1+0.95(t/T_{2,gg}^*)^2]^{-3/2}$ and $\kappa(t, T_{2,gg}^*)= -3 \arctan(0.97t/T_{2,gg}^*)$ \cite{S.Kuhr2005}, we obtain a coherence time $T_{2,gg}^* = 7.2 \pm 0.6 $ ms.
This coherence time is consistent with the inhomogeneous dephasing time $T_{2,gg}^* = 7.4$ ms deduced from the measured atom temperature at 5.2 $\mu$K, using formula $T_{2,gg}^* = 0.97\times2\hbar/\eta k_B T$ \cite{Kuhr2005} with $\eta = 3.85\times10^{-4}$. The residual decoherence results from the homogeneous dephasing, like the intensity fluctuation of dipole trap, magnetic fluctuation and so on.

\begin{figure}[htbp]
\centering
\includegraphics [width=0.48\textwidth]{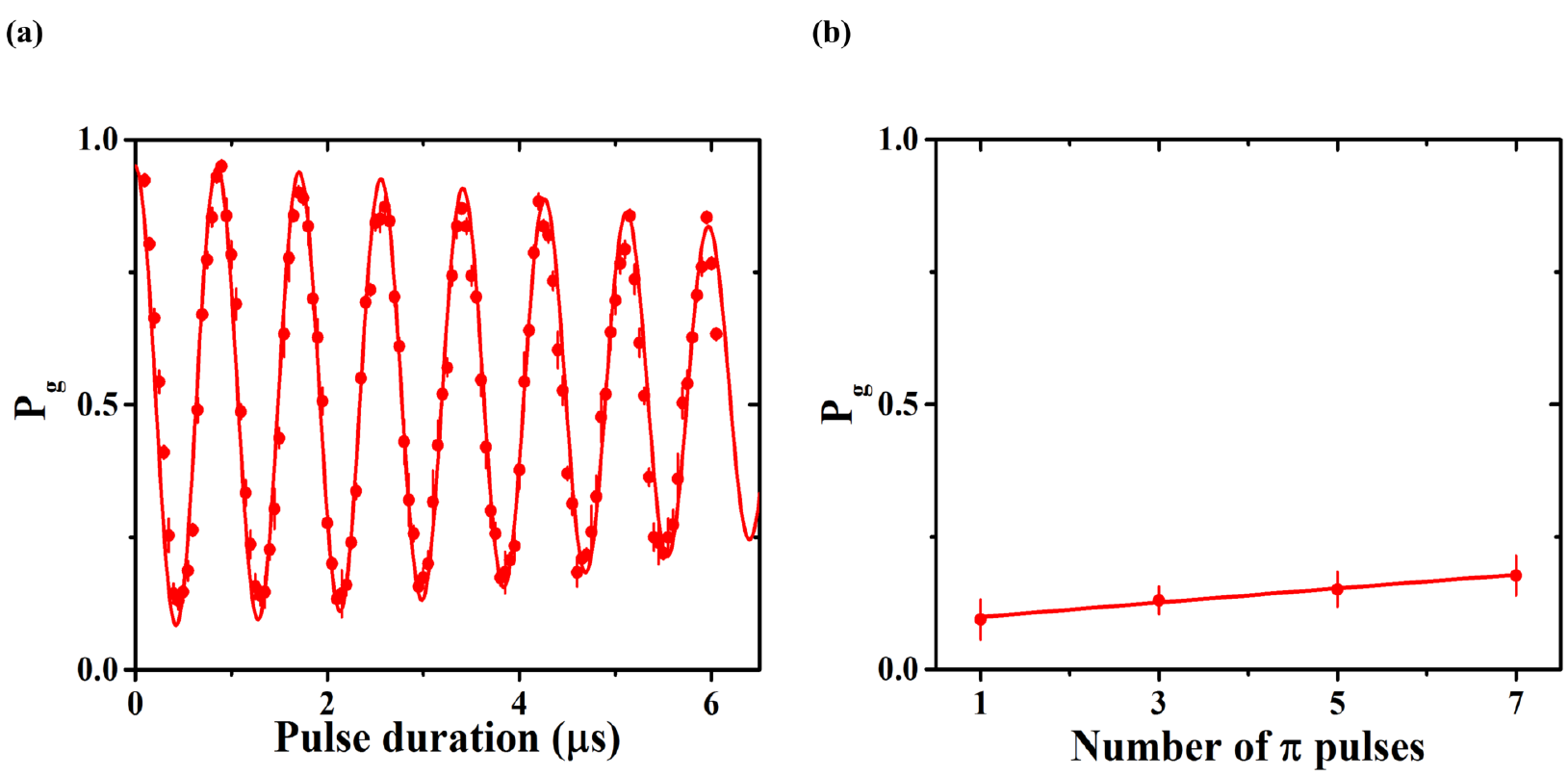}
\caption{(color online) (a) Rabi oscillation between $|1\rangle$ and $|r\rangle$ with the effective two-photon ground-Rydberg Rabi frequency $\Omega_0=2\pi \times 1.188$ MHz and the decay time of $14.4\pm1.4$ $\mu$s.  The probability in ground state is denoted as $P_g$ and the red dot is the experimental data. The red curve is a numerical simulation performed by Qutip \cite{Johansson2012QuTiP}. (b) Calibrating the detection efficiency $P_{rd}$ and Rydberg $\pi$-pulse excitation efficiency $P_{re}$ by applying an odd number of Rydberg $\pi$-pulses and detecting the survival probability in the ground state $|1\rangle$. By fitting the curve with $P = P_{1d} - P_{rd}P_{re}^n$, we obtain $P_{rd}$ of $88.7\%\pm0.5\%$ and $P_{re}$ of $98.4\%\pm0.4\%$.}
\label{fig:fig3}
\end{figure}

Next, we move on to discuss one of the core issues in this article, the Rydberg excitation and coherence between the ground state and the Rydberg state, which we will abbreviate as ground-Rydberg state for simplicity. The Rabi oscillation between $|1\rangle$ and $|r\rangle$ with the contrast of $0.44\pm0.01$ and decay time of $14.4\pm1.4$ $\mu$s is observed, as shown in Fig.\ref{fig:fig3}(a). The contrast of Rabi flopping is mainly limited by the detection efficiency of $|1\rangle$ (deonted as $P_{1d}$) and $|r\rangle$ (denoted as $P_{rd}$). According to our measurement, $P_{1d} = 97.2\% \pm 1.4\%$ depending on the trap-off time. $P_{rd}$ is calibrated by a sequence of odd Rydberg $\pi$ pulses and then detecting the survival probability, as shown in Fig.\ref{fig:fig3}(b), where we obtain $P_{rd} = 88.7\% \pm 0.5\%$ and the Rydberg $\pi$-pulse excitation efficiency $P_{re} = 98.4\pm0.4\%$. The $P_{rd}$ is limited by the finite Rydberg lifetime and the trap depth repulsing atoms in Rydberg state, and $P_{re}$ is mainly limited by the accuracy of $\pi$ pulse.
Besides, using the method of Ref. \cite{Zeng2018}, we estimate the single photon Rabi frequencies as $\Omega_r = 2\pi\times 215$ MHz for 780 nm laser and $\Omega_b = 2\pi\times 62$ MHz for 480 nm laser for this experiment.

The contributions to the decay of Rabi oscillation between $|1\rangle$ and $|r\rangle$ may come from the phase noise of the excitation lasers, laser intensity fluctuation, Doppler shift caused by the atom's residual thermal motion, off-resonant scattering from the intermediate level, finite Rydberg state lifetime and so on. In our setup, the original laser phase noise around 1 MHz is suppressed three orders by a high finesse cavity, and thus is practically negligible. Also, we ignore the effect of finite Rydberg lifetime, because the lifetime of $|r\rangle$ is about 209 $\mu$s which is much longer than excitation pulse time $\sim$ 6 $\mu$s. We build a four-level model for atoms in Qutip \cite{Johansson2012QuTiP} to take the rest factors into consideration and simulate the two-photon driving process, similar to method of Ref. \cite{S.deLeseleuc2018}. The off-resonant scattering from the intermediate level is included in this model by introducing a extra decay channel from $|p\rangle$ to another ground state $|g'\rangle$. The Doppler shift is added as a random detuning $\delta_D$ from a Gaussian distribution of width $2\pi \times 17$ kHz corresponding to atom temperature at 5.2 $\mu$K.
The laser intensity fluctuation comes from a combination of the instability of light intensity itself, $0.7\%$ in 780 nm laser and $0.7\%$ in 480 nm laser, and intensity fluctuation induced by the thermal distribution of cold atoms' motion. At 5.2 $\mu$K, the standard deviations of the position of the atom in the axial and radial directions of the trapping potential are $\Delta x_z = \sqrt{k_BT/m\omega_z^2} \simeq 2.2$ $\mu$m and $\Delta x_r = \sqrt{k_BT/m\omega_r^2} \simeq 0.3$ $\mu$m in our setup.
Accounting for the beam waist of 8 $\mu$m with the 480 nm laser and 10 $\mu$m with the 780 nm laser, the uncertainty of atom position results in intensity fluctuation of $2.2\%$ in the 480 nm laser and $1.8\%$ in the 780 nm laser. Taking these two factors into account, the total single-photon Rabi frequency deviation of 480 nm $\sigma_{480} = 1 \text{ MHz}$ and that of 780 nm $\sigma_{780} = 2.1 \text{ MHz}$ are also added into this simulation. The simulation result is indicated in Fig. \ref{fig:fig3}(a) as the red curve and the experimental results(red dot) is well fitted to the simulation result.
In order to realize better ground-Rydberg coherence, future improvements include larger laser spot, larger single-photon Rabi frequency of 480 nm laser and colder atoms.

\begin{figure}[htbp]
\centering
\includegraphics [width=0.48\textwidth]{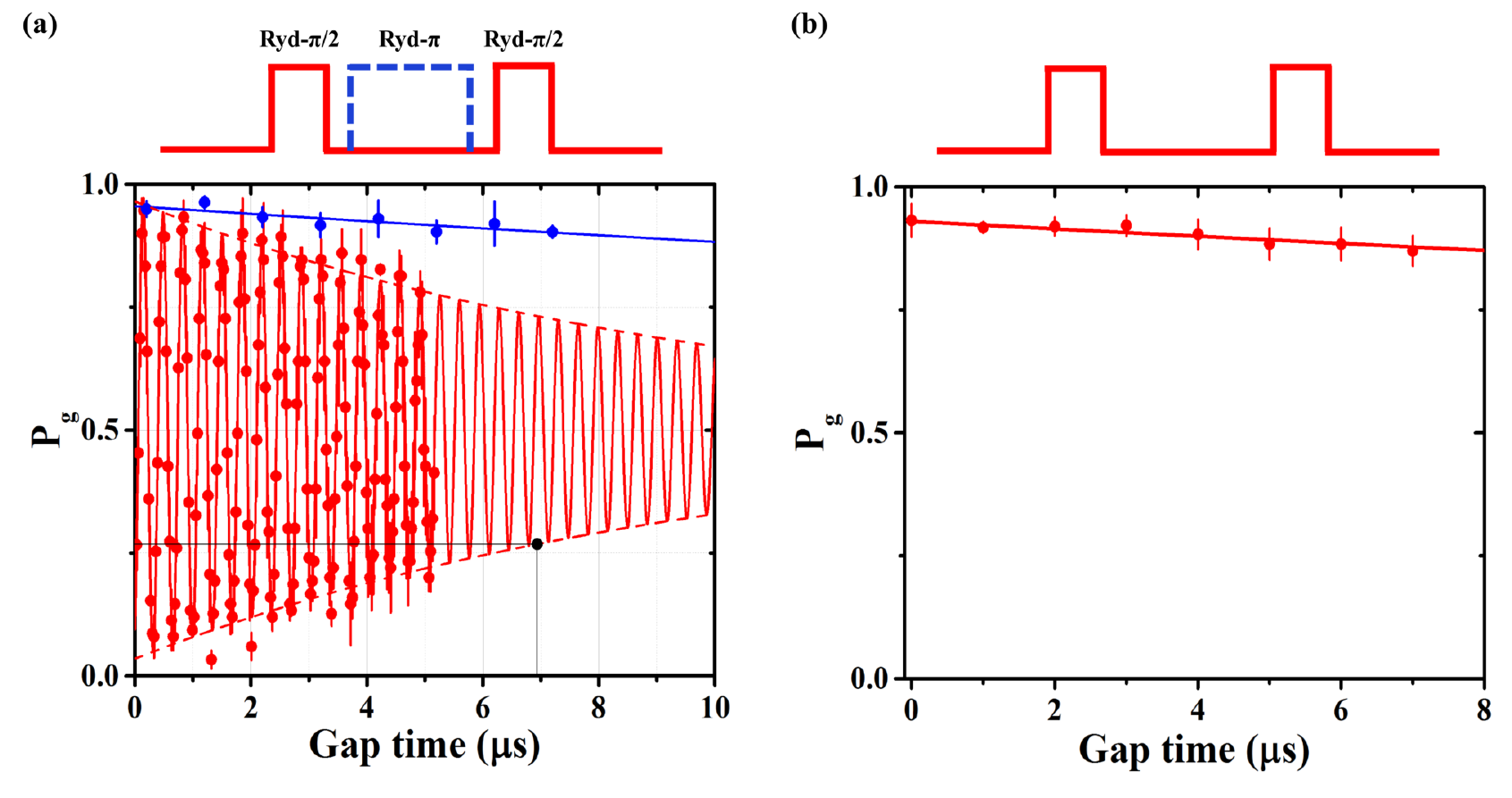}
\caption{(color online) (a) A ground-Rydberg Ramsey sequence is applied to measure $\tau_{gr}=10.0\pm0.9$ $\mu$s, where we fit the points to a damped sinusoidal curve (the blue dots). By inserting a Rydberg $\pi$-pulse (red) in the middle of the Ramsey sequence which acts as a spin-echo sequence, we cancel the effect of Doppler shift and obtain $ T_{2, \text{sp}}=57\pm14$ $\mu$s (offset is set to 0.5). (b) The lifetime of $|r\rangle$ is measured by exciting atoms from $|1\rangle$ to $|r\rangle$ with a Rydberg $\pi$ pulse and then de-exciting with a second Rydberg $\pi$ pulse after a gap time. The probability of the atom ending in ground state decays with an extracted lifetime of $T_1 = 122\pm25$ $\mu$s, which is fitted to an exponential decay model with no offset. }
\label{fig:fig4}
\end{figure}

The total coherence time $\tau_{gr}$ of ground-Rydberg state can be given by inhomogeneous dephasing time $T_{2,gr}^*$, homogeneous dephasing time $T_{2,gr}'$ and effective Rydberg lifetime $T_{1,r \rightarrow g}$:
\begin{equation}
\label{eq:T2}
\frac{1}{\tau_{gr}} = \frac{1}{T_{2,gr}^*} + \frac{1}{T_{2,gr}'} +\frac{1}{2T_{1,r\rightarrow g}}.
\end{equation}
In our setup, $\tau_{gr}$ is measured to be $10.0\pm0.9$ $\mu$s using a Ramsey sequence, as shown in Fig. \ref{fig:fig4}(a). Inhomogeneous dephasing $T_{2,gr}^*$ occurs because the atoms have different resonance frequencies due to Doppler shift, which can be reversed using a spin-echo technique by applying an additional $\pi$-pulse between the two Ramsey $\pi/2$-pulses  \cite{Kuhr2005}.  We carry out such a spin-echo measurement, and extract $T_{2, \text{sp}}$ of $57\pm14$ $\mu$s, as shown in Fig. \ref{fig:fig4}(b) (red curve). We assume that the spin-echo technique eliminates most of the inhomogeneous dephasing, so the decay time of spin-echo curve can be given by:
\begin{equation}
\label{eq:Tsp}
\frac{1}{T_{2, \text{sp}}} = \frac{1}{T_{2,gr}'} +\frac{1}{2T_{1,r\rightarrow g}}
\end{equation}
Thus, $T_{2,gr}^* = (1/T_{2,gr} - 1/T_{2, \text{sp}})^{-1} = 12.7$ $\mu$s, which is in good agreement with the Doppler effect induced dephasing time computed as $\sqrt{2}/k_{eff}\Delta v = 12.6$ $\mu$s \cite{Saffman_2011}.

To extract the homogeneous dephasing time $T_{2,gr}'$, we first measure $T_{1, r\rightarrow g}$ using two Rydberg $\pi$-pulses with different gap times. As shown in Fig. \ref{fig:fig4}(b), $T_{1,r\rightarrow g} \simeq T_1 = 122\pm25$ $\mu$s which is mainly limited by the effective lifetime of $|r\rangle$ of 209 $\mu$s at a 300 K ambient temperature \cite{I.I.Beterov2009} and the decay time induced by off-resonant scattering of 480 nm laser from $|e\rangle$ as $4\Delta^2/(\Omega_b^2 *\Gamma_e) = 940$ $\mu$s. From these two channels, we estimate $T_1 = 170$ $\mu$s, which shows some deviation from the experimental results. We think this deviation is caused by the measurement error when determining such long time on order of 170 $\mu$s. From Eq. \eqref{eq:Tsp}, we obtain $T_{2,gr}' = [1/T_{2, \text{sp}} - 1/(2T_1)]^{-1} = 74$ $\mu$s. This dephasing mainly results from laser intensity fluctuation, magnetic field fluctuation, residual laser phase noise and so on. Since $T_{2,gr}' = 74$ $\mu$s indicates a frequency deviation about $\sim2\pi \times 3$ kHz, we conclude that these factors which induce homogeneous dephasing are under well control in our setup.

\section{\label{sec:characterization the coherence of control qubit in a C-NOT gate}Coherence of control qubit in a CNOT gate}

One of the most important aspects of the CNOT gate is its ability to deterministically create entanglement \cite{PhysRevLett.81.3631}. For the neutral atom qubit platform, the standard method to prepare a Bell state based on the CNOT gate have been described in details in previous works \cite{L.Isenhower2010, Y.Zeng2017}. And the reported fidelity of entangled state is less than the corresponding values of CNOT gate truth table in relevant results \cite{K.Maller2015, Isenhower2010, Y.Zeng2017}. This is a slightly subtle issue, partly because the numbers of typical CNOT gate truth tables are concerned about state preparation efficiency and state flip efficiency, but not as quite informative about the coherence of control qubits and target qubits. Nevertheless, the production of entangled state based on CNOT gate involves linear superposition of states which strongly depends on the coherence properties of qubit atoms, as we will further explain.

\begin{figure}[htbp]
\includegraphics [width=0.48\textwidth]{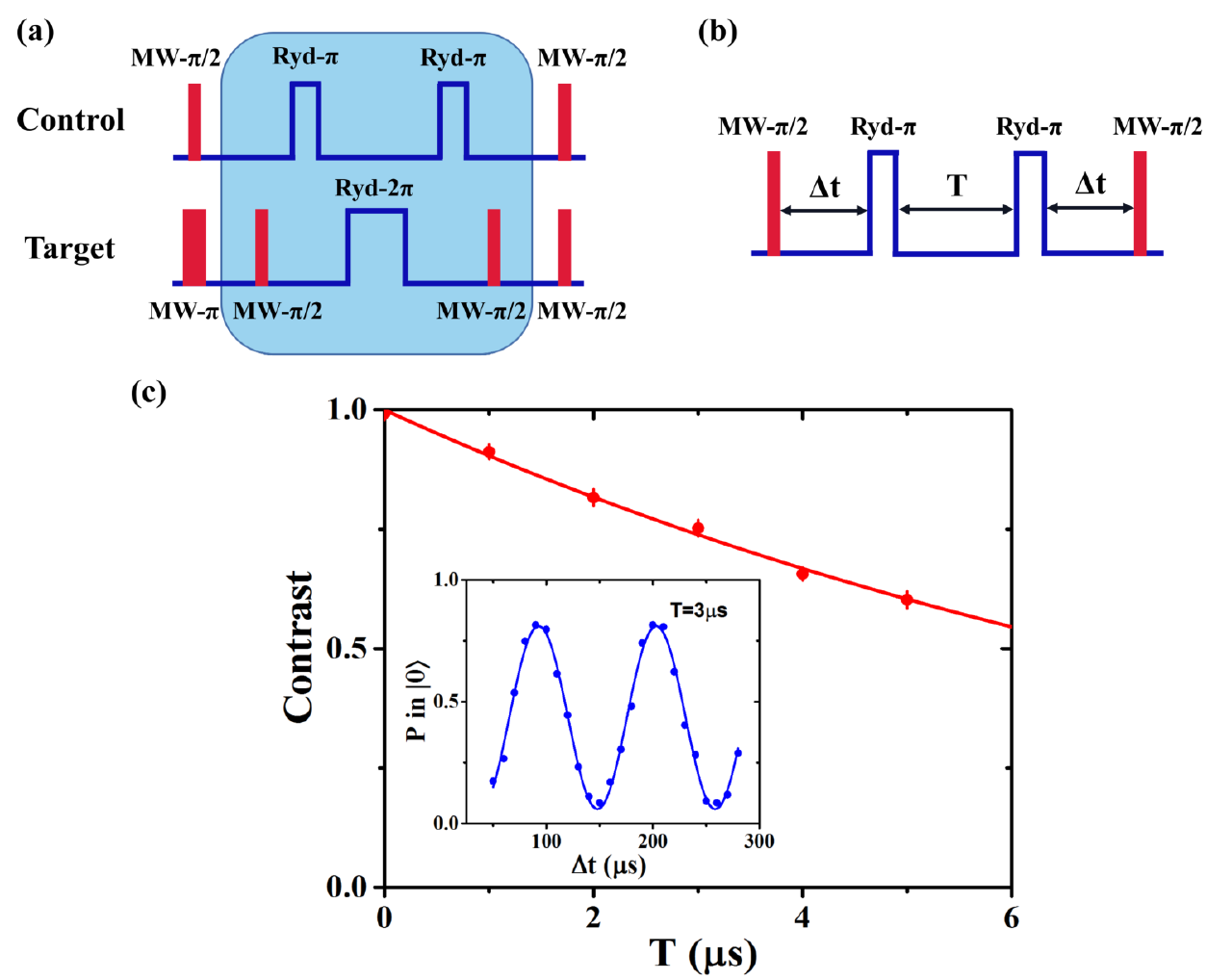}
\caption{(color online) Experiments to measure the coherence of control qubits. (a) The usual sequence of preparing a Bell state $(|1_C\rangle|1_T\rangle+|0_C\rangle|0_T\rangle)/\sqrt{2}$ using H-$C_Z$ type CNOT gate. The blue area is a uniform H-$C_Z$ type CNOT pulse sequence. GND-$\pi/2$ stands for $\pi/2$ pulse between $|0\rangle$ and $|1\rangle$. (b) The entanglement sequence applied on control qubit using the H-$C_Z$ type CNOT gate based on Rydberg blockade. In each experiment, we fix the value of $T$ and sweep the gap time $\Delta t$ to obtain a fringe. (c) The contrasts of fringes at $T=0, 1, 2, 3, 4, 5$ $\mu$s. They are fitted to a single exponential decay function with no offset and we get a fitted $1/e$ decay time of $T_\text{cont}=9.9\pm0.3$ $\mu$s. The inset shows the fringe at $T=3$ $\mu$s. }
\label{fig:fig5}
\end{figure}

Without loss of generality, the process to prepare the Bell state $(|1_C\rangle|1_T\rangle+|0_C\rangle|0_T\rangle)/\sqrt{2}$ is described below as a example, where the subscript '$C$' and '$T$' represent control qubits and target qubits respectively. As shown in our previous work\cite{Y.Zeng2017}, we generate the two-atom state $(|1_C\rangle+ i |0_C\rangle)|0_T\rangle/\sqrt{2}$ by applying a Raman $\pi/2$-pulse (GND-$\pi/2$) on control qubit; then, we apply a uniform C-NOT sequence (light blue shaded part in Fig. \ref{fig:fig3}(a)) on the two qubits to get the entanglement; finally, we apply two GND-$\pi/2$-pulse with a relative phase $\phi$ to verify the entanglement, as indicated in Fig. \ref{fig:fig5}(a).
During the entangling process, the target qubit only experiences a Rydberg $2\pi$-pulse whose net effect on target qubits is generating a $\pi$ phase while other pulses is only addressing the ground level. Therefore, the coherence of target qubit is determined by coherence between $|0\rangle, |1\rangle$.
We have experimentally recorded that the coherence time between $|0\rangle, |1\rangle$
is about $T_{2,gg}^*$ =7 ms in Sec. \ref{sec:Experiment setup},
so its influence can be neglected when comparing with CNOT gate time of several $\mu$s. However, the control qubit experiences a fundamentally different process. As indicated in Fig. \ref{fig:fig5}(b), the first GND $\pi/2$-pulse and the first Rydberg $\pi$-pulse bring the control qubit into a superposition state of ground and Rydberg levels $(|r_C\rangle+ i |0_C\rangle)/\sqrt{2}$; after a gap time T, a second such pulse sequence is applied in negative direction to recover the state of control qubit. By regarding the GND $\pi/2$-pulse and the Rydberg $\pi$-pulse as one unified step of operation, we observe that the process is in fact a Ramsey process of the ground-Rydberg state.

To verify this hypothesis, we characterize the coherence of the control qubit by measuring the fringe contrast when two Rydberg $\pi$-pulses are applied using the sequence in \ref{fig:fig5}(b).
In each measurement we keep the gap time $T$ constant between two Rydberg $\pi$-pulses and sweep the gap time $\Delta t$ between the neighboring Rydberg $\pi$-pulse and GND $\pi/2$-pulse, from which we get a Ramsey-type fringe. Such a sweep of the gap time  $\Delta t$ is repeated for the values of $T=0,1,2,3,4,5$ $\mu$s, as in Fig. \ref{fig:fig5}(b). Then, we get a series of Ramsey fringes and one of these fringes is shown in Fig. \ref{fig:fig5}(c) inset. We fit the contrasts of fringes to a single exponential decay function of $T$, and get a dephasing time of $T_{cont}=9.9 \pm 0.3$ $\mu$s which is well consistent with $\tau_{gr}=10.0 \pm 0.9$ $\mu$s, see in Fig. \ref{fig:fig5}(c).
Hence, in a CNOT gate based entanglement sequence, the control qubits experience a Ramsey process of ground-Rydberg state. And the dephasing time of control qubit in this process is in fact the ground-Rydberg dephasing time $\tau_{gr} = T_{cont}$ experimentally. In other words, ignoring errors from dephasing of target qubit which is mostly determined by the dephasing of ground level, the red curve in Fig. \ref{fig:fig5}(c) sets a high limit of fidelity of Bell state for a given value of $\tau_{gr}$.

\section{Theoretical Analysis}
\label{sec:theory}

Here, in order to provide a picture on the theoretical side, we discuss the results of numerical simulations for Doppler-induced dephasing caused by the thermal motion of atoms, with respect to the phenomena in the ground-Rydberg Ramsey process and the two-qubit gate fidelity.

\begin{figure}[h]
\centering
\includegraphics [width=0.48\textwidth]{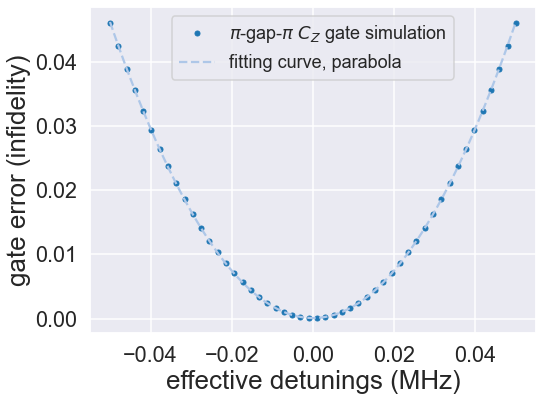}
\caption{(color online) Numerical simulation for the relation between the $C_Z$ gate error and the effective detuning of the ground-Rydberg driving laser on the control qubit atom, with respect to a single-photon ground-Rydberg transition. The total gate time is set as 2 $\mu$s, and the $\pi$-pulse is set as a square pulse with duration 0.5 $\mu$s. The Rydberg blockade strength is set as $2\pi \times 1000$ MHz. The gate error (infidelity) is defined as $\mathcal{E} = 1 - F$ with $F$ being the fidelity.}
 \label{fig:infidelity_vs_detuning}
\end{figure}

To begin with, we evaluate how the non-zero velocity of the qubit atoms can adversely influence the $C_Z$ gate fidelity. In terms of atom-light interaction, this reduces to an effective detuning of the driving laser, and we carry out such a numerical simulation for Rb atom with single-photon ground-Rydberg transition. A typical result is shown in Fig. \ref{fig:infidelity_vs_detuning}, with a fitting to the parabola curve. The Rabi frequency of the laser driving the ground-Rydberg transition is set at 1 MHz.

\begin{figure}[bh]
\centering
\includegraphics [width=0.48\textwidth]{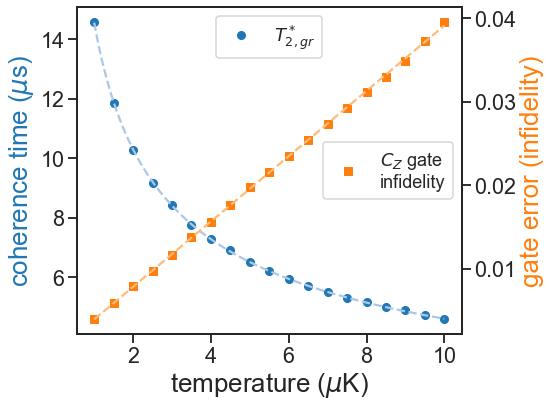}
\caption{(color online) Numerical simulation for the decoherence effects caused by the finite temperature of Rb cold atoms, for single-photon ground-Rydberg transition. Simulation of $T_{2,gr}^*$ is expressed in blue dots with fitting curve in the form of $f(x) = 1/\sqrt{x}$, and simulation of $C_Z$ gate fidelity is expressed in squares with linear fitting curve.}
 \label{fig:temperature_effects}
\end{figure}

Consequently, we proceed to study the effects of the residual thermal motion of cold atoms, in terms of temperature vs. the intrinsic inhomogeneous ground-Rydberg coherence time $T_{2,gr}^*$ and temperature vs. $C_Z$ gate fidelity. We carry out numerical simulations according to the Monte-Carlo method, where we assume that the velocity distribution of cold atoms obey the 3D Maxwell-Boltzmann distribution. A typical result is shown in Fig. \ref{fig:temperature_effects}. The information of $T_{2,gr}^*$ is extracted from the typical Ramsey sequence with driving laser's Rabi frequency set as 1 MHz, and the $C_Z$ gate pulse sequence is set the same as Fig. \ref{fig:infidelity_vs_detuning}.

Although Fig. \ref{fig:temperature_effects} only considers the intrinsic decoherence source of the residual thermal motion of cold atoms, we can deduce the relation between $T_{2,gr}^*$ and $C_Z$ gate error $\mathcal{E}$ when $\mathcal{E} \ll 1$. Then, with respect to the $\pi$-gap-$\pi$ type $C_Z$ gate in terms of 2 $\mu$s gate time as that of Fig. \ref{fig:infidelity_vs_detuning}, we arrive at the following relation:
\begin{equation}
\label{eq:fidelity_vs_T2*}
\mathcal{E} \approx 0.836  \times \big(\frac{1}{T_{2,gr}^*}\big)^2,
\end{equation}
where the term of $T_{2,gr}^*$ is in the unit of $\mu$s. Certainly, many other decoherence factors exist and affects the $T_{2,gr}^*$ and two-qubit gate fidelity, from both technical and intrinsic sources. Therefore, extending the applicability Eq. \eqref{eq:fidelity_vs_T2*} to the general sense does not necessarily guarantee a satisfying description. Nevertheless, it can serve as a reasonable reference for the fundamental relation between the single-atom ground-Rydberg coherence property and the two-qubit Rydbert blockade gate performance.

According to the latest research developments \cite{K.Maller2015, S.deLeseleuc2018, H.Levine2019, T.M.Graham2019}, the practical fidelity of two-qubit gate is limited by two major factors assuming that the Rydberg blockade strength is adequate.

The first one is the SPAM error, which includes atom preparation error in the trap, optical pumping error and state measurement error. The atom preparation error results from background collisions at finite vacuum pressure which can be reduced by faster measurements and improved vacuum conditions. Measuring the ratio of pumping time to depumping time is an effective method to measure the state preparation infidelity and the fidelity of optical pumping is usually $>$99\%. Experimentally, it is difficult to distinguish the optical pumping error from the atom preparation error, both of which contribute to the reduction of $|0\rangle$ and $|1\rangle$'s Rabi oscillation contrast. Generally, the primary technique of state read-out is to apply a resonant laser pulse to push out the atoms in $|1\rangle$ (in F=2 for $^{87}$Rb generally) from the trap. This method correctly identifies atoms in $|0\rangle$ (in F=1 for $^{87}$Rb generally), but it can mistakenly count in atoms loss through background loss processes or by residual Rydberg excitation for atoms in $|1\rangle$, resulting in an overestimation of the population in $|1\rangle$ \cite{H.Levine2019}. Therefore, for any measurements involving Rydberg excitation, it is better to collect measurement statistics both with and without the push-out pulse \cite{H.Levine2019, T.M.Graham2019}. Such a practice provides an upper bound on how much leakage out of the qubit subspace has occurred, and therefore also gives a lower bound on the true population in $|1\rangle$. Besides, nondestructive and faster readout methods have been demonstrated by several groups and these methods can distinguish the position of atoms in the same time \cite{MinhoKwon2017, M.Martinez-Dorantes2017}. These new ideas can contribute to improvement of fidelity of two qubit gate in near future.

The second main factor which handicapped the improvement of fidelity is imperfections in ground-Rydberg coherence. One of the consequence is the imperfect Rydberg excitation, which have been analyzed in detail in Ref. \cite{S.deLeseleuc2018}. It can result from Doppler-induced dephasing effect, spontaneous emission from the intermediate levels for two-photon transition, laser phase noise, finite lifetime of the Rydberg state, uniformity of excitation beam, stray electric fields and the fluctuations of magnetic fields. Reducing technical limitations has become a focus recently, such as reducing laser phase noise by spectral filter \cite{H.Levine2018} or by adjusting locking parameters carefully \cite{T.M.Graham2019}, shorter pulse time and more uniform spot. In the future, research efforts are anticipated to mitigate the stray electric fields and the fluctuations of magnetic fields. In particular, due to the finite temperature of atoms, Doppler-induced dephasing is an intrinsic source of error \cite{T.Wilk2010}. Very often, Rydberg excitation is realized by the two-photon transition and the two lasers' wave vectors cannot ideally cancel out, resulting in an effective wave vector mismatch $k_{eff}$. In addition, atoms with finite temperature results in a one-dimensional rms velocity spread of $\Delta v$. In other words, this is equivalent to a spread of effective detunings on the order of $k_{eff} \Delta v$.

Last but not least, residual thermal motion of cold atoms is not the only source contributing to $T_{2,gr}^*$ of the ground-Rydberg transition. Experimentally, the coherence purely due to Doppler-induced dephasing $T_{2D}^*$ is larger than $T_{2,gr}^*$. Virtually, all sources limiting the two-qubit gate fidelity also contribute to the limiting of $T_{2,gr}^*$, except for the Rydberg blockade strength.

\begin{figure}[h]
\centering
\includegraphics [width=0.48\textwidth]{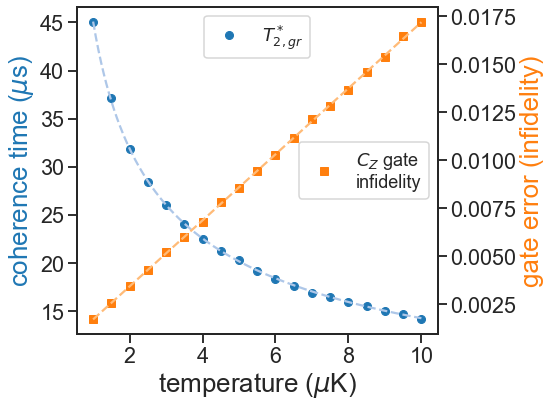}
\caption{(color online) Numerical simulation for the decoherence effects caused by the finite temperature of Rb cold atoms, for two-photon ground-Rydberg transitions. Rabi frequencies are set as $\Omega_r = 2\pi\times 215$ MHz for 780 nm laser and $\Omega_b = 2\pi\times 62$ MHz for 480 nm laser respectively, while the one-photon detuning is set as $\Delta = 2\pi\times -5.7$ GHz. Spontaneous emissions from the intermediate levels are not considered in this simulation.}
 \label{fig:temperature_effects_2photon}
\end{figure}

On the other hand, we have also carried out numerical simulations for the two-photon ground-Rydberg transitions, according to the experimental conditions laid out in Section \ref{sec:Experiment setup} and \ref{sec:Coherent control and coherence of single qubit}. We note that the experimentally observed $T_{2,gr}^*$ is smaller than the computed values in Fig. \ref{fig:temperature_effects_2photon}, and this is a direct consequence of the fact that more sources are contributing to the decoherence effects of the ground-Rydberg transition experimentally, besides the residual thermal motion of cold atoms.

\section{Conclusion and Outlook}
\label{sec:Discussion}

We have analyzed the underlying connection between the inhomogeneous dephasing time $T_2^*$ of the ground-Rydberg transition and the performance of the Rydberg blockade two-qubit gate, both experimentally and theoretically. Our results have clearly demonstrated the importance of the ground-Rydberg coherence in characterizing the neutral atom two-qubit gates. In other words, improving $T_2^*$ serves as a prerequisite to enhance the entangling gate performances.

Based on our discussions so far, we'd like to provide an outlook into the future on how to achieve better fidelities. The first method is to use advanced atom cooling techniques such as Raman sideband cooling to get cooler atoms to suppress the Doppler-induced dephasing \cite{Madjarov2020, Lorenz2020}. The second method is to decrease the ground-Rydberg excitation pulse time, if higher laser power can be obtained. The third one, also the most promising method as it seems to us, is to explore a new two-qubit gate scheme rather than the $\pi$-gap-$\pi$ protocol, in order to get a more friendly requirement of ground-Rydberg coherence time $T_2^*$. There have been many encouraging progress to develop new scheme for neutral atoms two-qubit gates using Rydberg blockade recently \cite{Theis2016, Sun2020, Saffman2020, Mitra2020, Zhang2020}. Some of them contain interesting features such as avoiding shelving population in the Rydberg states and completing $C_Z$ gate within a single pulse, and we are looking forward to the experimental performance.

\begin{acknowledgments}
This work was supported by the National Key R\&D Program of China (under Grant Nos.  2016YFA0302800 and 2016YFA0301504), the National Natural Science Foundation of China under Grant Nos. U20A2074 and 12074391, the Strategic Priority Research Program of CAS (grant XDB 21010100), the Youth Innovation Promotion Association of CAS (grants 2017378) and K.C.Wong Education Foundation (GJTD-2019-15).
\end{acknowledgments}

\bibliography{RydbergT2star_ref}

\end{document}